\documentclass[twocolumn,pra,superscriptaddress,showpacs,aps,floatfix]{revtex4-1}
\usepackage{color}
\usepackage{amsmath}
\usepackage{amssymb}
\usepackage{txfonts}
\usepackage{graphicx}
\usepackage{epstopdf}
\usepackage[unicode]{hyperref}
\hypersetup{pdftitle={2D SOC BEC},
 pdfauthor={},
 pdfsubject={2D SOC BEC},
  colorlinks=true,
    linkcolor=red,
    citecolor=magenta,
    filecolor=black,
    urlcolor=blue}


\begin{document}

\title{Theoretical Origin of CP violation in the FCNC-free 2HDM  \\
and Its Extensions to the Standard Model and 3HDM}

\author{Chilong Lin}
\email{lingo@mail.nmns.edu.tw}

\affiliation{National Museum of Natural Science, 1st, Guan Chien RD., Taichung, 40453 Taiwan, ROC}

\date{Version of \today. }

\begin{abstract}
In this manuscript, the general FCNC-free and CP-violating pattern of quark-mass matrices in the 2HDM derived in our previous investigations is revised.
This revised pattern is to be diagonalized analytically with no symmetries imposed.
The unitary transformation matrices thus derived depend on only two parameters in each quark type and subsequently lead to a CKM matrix which depends on at most four parameters.
The fitting of theoretically derived CKM elements and their corresponding empirical values are as good as ${\bf O}(10^{-2})$ at tree-level.
In a phenomenological way which imposes several fine-tuning parameters into the CKM matrix suitably, the fitting is further improved to ${\bf O}(10^{-4})$.
After the derivation, we find this hypothesis also applies to the Standard Model and even a model with three Higgs doublets.
This will be a big progress in the derivation of a theoretical origin of CP violation.

\end{abstract}
\maketitle


\section{Introduction}

In our previous investigations ~\cite{Lin1988, Lee1990, Lin2013}, four Flavor-Changing-Neutral Current (FCNC)-free patterns of quark-mass matrices in the Two-Higgs-Doublet Models (2HDMs) had been derived analytically with a special Hermitian condition firstly proposed in ~\cite{Branco1985}.  \\

In the first of such FCNC-free 2HDMs ~\cite{Lin1988}, a $S_3$ symmetry was imposed among the three fermion generations.
However, it solved only the FCNC problem in a 2HDM but not the problem of theoretical origin of CP violation.
In a subsequent investigation following the same concept ~\cite{Lin2013}, three additional such FCNC-free matrix pairs were discovered with residual $S_2$ symmetries between two of the three generations.
The $S_3$-symmetric pattern together with these three $S_2$-symmetric patterns provide us a way to derive a complex Cabbibo-Kobayashi-Maskawa (CKM) matrix and thus breaks CP symmetry explicitly.
But, the CKM elements thus derived have very different amplitudes to corresponding experimentally detected values.
That drives us to a more general investigation without any symmetries and which is the subject to be studied in this manuscript.  \\

In \cite{Lin2013}, a general pattern for FCNC-free matrix pairs was derived by assuming the mass matrices are Hermitian of themselves.
The Hermitian assumption reduced the number of parameters in a quark type from eighteen down to nine.
Besides, it also led to a special Hermitian condition $M_1 M_2^{\dagger}-M_2 M_1^{\dagger} =0$ which provided us four extra conditions among the parameters and further  reduced the number of them down to five.
In that manuscript we did not diagonalize the mass matrices directly to achieve analytical solutions since it seemed very difficult at that time.
There, an assumption among $A$ parameters, $A=A_1=A_2=A_3$, were employed to simplify the pattern so as to achieve matrix pairs satisfying the Hermitian condition mentioned above.
The matrix pairs thus derived were found to possess residual $S_2$ symmetries between two of the three generations.
That gave us totally four FCNC-free patterns and thus four corresponding $U$ matrices so as to lead to several complex CKM matrices.
But the derived CKM matrices do not fit the empirical values very well at that time. \\

However, if we remove the assumptions among $A$ parameters and diagonalize the mass matrices revised directly,
analytically derived mass eigenvalues and the corresponding $U$ matrix are in fact achievable as to be shown  in section II.
It is amazing the general pattern of $U$ matrices thus derived depends only on two of the five parameters in a quark type.
That indicates the CKM matrix thus derived depends on at most four parameters.
When fitting the derived CKM elements with corresponding empirical values, the largest deviation between them is only about ${\bf O}(10^{-2})$ which is orders better than those derived in ~\cite{Lin2013}. \\

In order to improve the fitting between the derived CKM elements and the empirical values further,
some phenomenological fine-tuning parameters are imposed as to be shown in section III.
In this way, the fittings are further improved to ${\bf O}(10^{-4})$ which is as good as the experimental errors at present.
However, the physical meaning of these fine-tunings is still obscure.
That surely indicate the need of more efforts in our future researches.  \\

After the derivation, we find this hypothesis not only applies to the 2HDM but also to the Standard Model and a model with three Higgs doublets.
The applications of it on the Standard Model and the Three-Higgs-Doublet Model (3HDM) will be discussed in section IV.
The section V will be devoted to the conclusions and discussions on the direction of our future investigations. \\

\section{The General Pattern of FCNC-free Mass Matrices}

As it was well known, CP symmetry can only be violated in the standard model (SM) of eletroweak interactions by ranking the Yukawa couplings between fermions and Higgs fields suitably to achieve complex CKM elements explicitly.
However, for decades, no one knows how these couplings, or equivalently the elements of fermion-mass matrices,
should be ranked to achieve such a matrix pattern since there are too many parameters in them to be diagonalized analytically.
That induced the extension of SM with an extra Higgs doublet to break CP symmetry spontaneously \cite{TDLee1973}.
But, the extra Higgs doublet not only failed to give an origin of CP violation, but also brought in an extra FCNC problem which arises when those two components $M_1$ and $M_2$, which correspond to Higgs doublets $\Phi_1$ and $\Phi_2$ respectively, of a fermion mass matrix $M=M_1+M_2$ are not diagonalized simultaneously.  \\

However, the FCNC problem is theoretically solvable if one can find a matrix pair with which both $M_1$ and $M_2$ were diagonalized simultaneously.
In our previous investigations \cite{Lin1988, Lee1990}, a $S_3$ Symmetry among three fermion generations was imposed to achieve the first such FCNC-free matrix pair.
Subsequently, with an interesting condition between a pair of Hermitian matrices, a more general pattern of such FCNC-free matrix pairs were derived \cite{Lin2013} and the $S_3$-symmetric matrix pair was found to be included in it as a special case.
In that manuscript, an assumption $A=A_1=A_2=A_3$ was imposed to simplify the matrix pattern and thus three more FCNC-free matrix pairs were achieved so as to bring in several complex CKM matrices.
However, the assumption imposed there is in fact unnecessary since the matrix pair demonstrated there is analytically solvable. \\

In what follows, no symmetry will be imposed and the general pattern of such FCNC-free matrix pairs will be diagonalized analytically.
Here, we will start the derivation with adoption of Eq.(5) in \cite{Lin2013} as
\begin{eqnarray}
M_1 = \left( \begin{array}{ccc} A_1 &  B_1 &  B_2  \\  B_1 &  A_2 &  B_3  \\ B_2 & B_3 & A_3 \end{array}\right) ,  ~~~~~
M_2 = i \left( \begin{array}{ccc} 0 & C_1 & C_2   \\  -C_1 & 0 & C_3  \\ -C_2 & -C_3 & 0 \end{array}\right).
\end{eqnarray}

The Hermitian condition $M_1 M_2^{\dagger}-M_2 M_1^{\dagger} =0$ gives us four equations
\begin{eqnarray}
B_1 C_1 &=& ~-B_2 C_2 = ~~B_3 C_3, \\
(A_1-A_2) &=& ~~~(B_3 C_2+B_2 C_3)/ C_1,  \\
(A_3-A_1) &=& ~~~(B_1 C_3-B_3 C_1)/ C_2,  \\
(A_2 -A_3) &=& -(B_2 C_1+B_1 C_2)/ C_3,
\end{eqnarray}
which are Eq.(9)-(12) in \cite{Lin2013} and will be used to reduce the number of parameters down to five in the followings. \\

With these four equations, one may replace four of the parameters with the others.
Here we choose to keep $A_3$, $B_3$, $C_3$, $B_1$ and $B_2$ while $A_1$, $A_2$, $C_1$ and $C_2$ are to be replaced by
\begin{eqnarray}
A_1 &=& A_3 +B_2 (B_1^2 -B_3^2)/ B_1 B_3,\nonumber \\
A_2 &=& A_3 +B_3 (B_1^2 -B_2^2)/ B_1 B_2, \nonumber \\
C_1 &=& B_3 C_3 /B_1,~~~~~~
C_2 = -B_3 C_3 /B_2 .
\end{eqnarray}

The mass matrix $M$ now becomes
\begin{eqnarray}
M &=& M_1 +i M_2 \nonumber \\
&=& \left( \begin{array}{ccc}
A + x B (y- {1 \over y})     & y B    & x B   \\
y B    & A +B ({y \over x}-{x \over y})  &  B \\
x B    & B                & A   \end{array}\right) +
 i \left( \begin{array}{ccc}
0   & {C \over y} & -{C \over x}   \\
-{C \over y} & 0  & C \\
 {C \over x} & C  & 0   \end{array}\right) \nonumber \\ 
\end{eqnarray}
if we let $A \equiv A_3$, $B \equiv B_3$, $C \equiv C_3$ and $x \equiv B_2 / B_3$, $y \equiv B_1 / B_3$. \\

Diagonalizing Eq.(7) analytically and directly, the mass eigenvalues will be given as
\begin{eqnarray}
M^{dia.} &=& (m_1,~ m_2,~m_3) = (X-Y,~ X+Y,~ Z),
\end{eqnarray}
where $X=A-{x \over y}B,~Y={\sqrt{x^2 +y^2 +x^2 y^2} \over {x y}}C$ and $Z=A+{{(x^2+1) y} \over x}B$. \\

The eigenvectors or the $U$ matrix which diagonalize Eq.(7) are given as
\begin{eqnarray}
&U& = \nonumber \\
& &\left( \begin{array}{ccc}
{-\sqrt{x^2+y^2} \over \sqrt{2(x^2+y^2+x^2 y^2)}} &~ {{x(y^2-i \sqrt{x^2+y^2+x^2 y^2})} \over {\sqrt{2} \sqrt{x^2+y^2} \sqrt{x^2+y^2+x^2 y^2}}} &~{{y(x^2+i \sqrt{x^2+y^2+x^2 y^2})} \over {\sqrt{2} \sqrt{x^2+y^2} \sqrt{x^2+y^2+x^2 y^2}}}  \\
 {-\sqrt{x^2+y^2} \over \sqrt{2(x^2+y^2+x^2 y^2)}} &~ {{x(y^2+i \sqrt{x^2+y^2+x^2 y^2})} \over {\sqrt{2} \sqrt{x^2+y^2} \sqrt{x^2+y^2+x^2 y^2}}} &~{{y(x^2-i \sqrt{x^2+y^2+x^2 y^2})} \over {\sqrt{2} \sqrt{x^2+y^2} \sqrt{x^2+y^2+x^2 y^2}}} \\
 {{x y} \over \sqrt{x^2+y^2+x^2 y^2}} &~ {y \over \sqrt{x^2+y^2+x^2 y^2}} &~{x \over \sqrt{x^2+y^2+x^2 y^2}} \end{array}\right).
\end{eqnarray}
It is amazing that all elements of $U$ are completely independent of three of those five parameters.
They depend on the parameters $x$ and $y$ only! \\

With Eq.(9), it's always possible for us to have $U^{(u)}$ for up-type quarks different from $U^{(d)}$ for down-type quarks if the parameters $x$ and $y$ have different values in different quark types.
The CKM matrix thus derived can be given with the parameters in $U^{(d)}$ assigned to primed ones $x'$ and $y'$ as
\begin{eqnarray}
V_{CKM} = \left( \begin{array}{ccc} V_{ud} & V_{us} & V_{ub} \\ V_{cd} & V_{cs} & V_{cb} \\ V_{td} & V_{ts} & V_{tb} \end{array}\right)
= \left( \begin{array}{ccc}
r e^{-i \delta_1} & s e^{i \delta_2} & p e^{-i \delta_3} \\
s e^{-i \delta_2} & r e^{i \delta_1} & p e^{i \delta_3} \\
p e^{i \delta_4} & p e^{-i \delta_4} & q \end{array}\right),
\end{eqnarray}
where
\begin{widetext}
\begin{eqnarray}
 V_{ud} &=& V_{cs}^* = r~ e^{-i \delta_1} = {{(x^2+y^2)(x'^2+y'^2)+(x x' +y y')(x y x' y' +\sqrt{x^2+y^2+x^2 y^2} \sqrt{x'^2+ y'^2 +x'^2 y'^2})}
   \over {2 \sqrt{ x^2+y^2} \sqrt{ x'^2+y'^2}\sqrt{x^2+y^2+x^2 y^2} \sqrt{x'^2+y'^2+x'^2 y'^2} }} \nonumber \\
   &+& i~ {{ (x y' -x' y)(x' y' \sqrt{x^2+y^2+x^2 y^2} +x y \sqrt{x'^2 +y'^2 +x'^2 y'^2})}
   \over {2 \sqrt{ x^2+y^2} \sqrt{ x'^2+y'^2}\sqrt{x^2+y^2+x^2 y^2}\sqrt{x'^2+y'^2+x'^2 y'^2}}}, \\
V_{us} &=& V_{cd}^* = s~ e^{i \delta_2} = {{(x^2+y^2)(x'^2+y'^2)+(x x' +y y')(x y x' y' -\sqrt{x^2+y^2+x^2 y^2} \sqrt{x'^2+ y'^2 +x'^2 y'^2})}
   \over {2 \sqrt{ x^2+y^2} \sqrt{ x'^2+y'^2}\sqrt{x^2+y^2+x^2 y^2} \sqrt{x'^2+y'^2+x'^2 y'^2} }} \nonumber \\
   &+& i~ {{ (x y' -x' y)(x' y' \sqrt{x^2+y^2+x^2 y^2} -x y \sqrt{x'^2 +y'^2 +x'^2 y'^2})}
   \over {2 \sqrt{ x^2+y^2} \sqrt{ x'^2+y'^2}\sqrt{x^2+y^2+x^2 y^2}\sqrt{x'^2+y'^2+x'^2 y'^2}}}, \\
V_{ub} &=& V_{cb}^* = p~ e^{-i \delta_3} =  {{[y' y^2(x-x')+ x' x^2 (y-y')]+ i (x y' -x' y) \sqrt{x^2+y^2+x^2 y^2}}  \over {2 \sqrt{ x^2+y^2} \sqrt{x^2+y^2+x^2 y^2}\sqrt{x'^2+y'^2+x'^2 y'^2}}}, \\
  V_{td} &=& V_{ts}^* = p~ e^{i \delta_4} = {{[y y'^2 (x'-x)+ x x'^2 (y'-y)]+  i (x y' -x' y) \sqrt{x'^2+y'^2+x'^2 y'^2}} \over {2 \sqrt{ x^2+y^2+x^2 y^2} \sqrt{ x'^2+y'^2}\sqrt{x'^2+y'^2+x'^2 y'^2}} }, \\
 V_{tb} &=& q = {{x x'+y y' +x y x' y'}\over {\sqrt{ x^2+y^2+x^2 y^2} \sqrt{x'^2+y'^2+x'^2 y'^2}}}.
\end{eqnarray}
\end{widetext}

The parameters $p$, $q$, $r$ and $s$ are amplitudes of corresponding CKM elements and $\delta_1$, $\delta_2$, $\delta_3$ and $\delta_4$ are phases.
It is interesting that amplitudes $\vert V_{ub} \vert = \vert V_{cb}\vert = \vert V_{td} \vert = \vert V_{ts} \vert = p$ predicted in Eq.(10) do not match the empirical values obviously. \\

If we compare the derived CKM elements with empirical values in ~\cite{Patrignani2016}
\begin{widetext}
\begin{eqnarray}
V_{CKM}^{empirical}=\left( \begin{array}{ccc}
V_{ud} & V_{us} & V_{ub} \\ V_{cd} & V_{cs} & V_{cb} \\ V_{td} & V_{ts} & V_{tb} \end{array}\right)
 =\left( \begin{array}{ccc}
 0.97434_{-0.00012}^{+0.00011} & 0.22506 {\pm 0.00050} & 0.00357 {\pm 0.00015} \\
 0.22492 {\pm 0.00050} & 0.97351 {\pm 0.00013} & 0.0411 {\pm 0.0013} \\
 0.00875_{-0.00033}^{+0.00032} & 0.0403 {\pm 0.0013} & 0.99915 {\pm 0.00005}
 \end{array}\right),
\end{eqnarray}
\end{widetext}
we will find some of the equations do not match Eq.(16) very well at tree level. \\

For example, the theoretically predicted $V_{us}=V_{cd}^*$ in Eq.(12) fit the empirical $\vert V_{us} \vert = 0.22506\pm 0.00050$ and $\vert V_{cd} \vert = 0.22492\pm 0.00050$ very well.
It indicates that $0.22542$ $ \geq ~\vert V_{us} \vert=\vert V_{cd} \vert ~\geq 0.22456$.
However, the theoretically predicted $\vert V_{ud} \vert =\vert V_{cs} \vert =r$ in Eq.(11) does not fit the empirical values $\vert V_{ud} \vert=0.97434^{+0.00011}_{-0.00012}$ and $\vert V_{cs} \vert=0.97351 \pm 0.00013$ so well since an overlapping zone between them like the one in Eq.(12) is absent.
But, the difference between the lower bound of $\vert V_{ud} \vert $ and the upper bound of $\vert V_{cs} \vert$ is only about 0.00058 which is so small that we may attribute it as loop corrections rationally.
However, the equations (13) and (14) are far beyond our expectation since they predict $\vert V_{ub} \vert =\vert V_{cb} \vert=\vert V_{td} \vert=\vert V_{ts} \vert =p$ which are very different from the empirical values $\vert V_{cb} \vert / \vert V_{ub} \vert = 11.5$ and $\vert V_{ts} \vert / \vert V_{td} \vert = 4.61$.
Such predictions deviate from the empirical values by an amount as large as ${\bf O}(10^{-2})$ or about 0.01877. \\

However, the CKM elements derived in this section are already orders improved when compared to those derived in ~\cite{Lin2013}.
For example, $\vert V_{ub} \vert$=2/3 was predicted in ~\cite{Lin2013}, which is 187 times that of the empirical value $\vert V_{ub} \vert^{emp.} =0.00357$ in Eq.(16).
While in this section, the best fit of $\vert V_{ub} \vert$ is achieved by assuming $p= {{\vert V_{ub} \vert^{emp.} +\vert V_{cb} \vert^{emp.}}\over 2}= 0.02234$, where the super-index "emp." indicates they are empirical values in Eq.(16).
The deviations of such a $p$ value from $\vert V_{ub} \vert^{emp.}$ and $\vert V_{cb} \vert^{emp.}$ are both 0.01877, which is only about 5.26 times that of $\vert V_{ub} \vert^{emp.}$.
Obviously, the predictions are improved orders better than those in ~\cite{Lin2013}.
\\

\section{A Phenomenological Improvement of CKM Matrix}

As mentioned above, some theoretically derived CKM elements do not fit the empirical values perfectly at tree-level.
However, the largest difference between them is only about 0.01877 for $\vert V_{ub} \vert$ and $\vert V_{cb} \vert$ if $p=0.02234$.
Thus, it is rational to consider such tiny deviations as contributions from loop corrections.
However, at this stage we will not do such loop calculations directly.
Instead, we would like to employ some fine-tuning parameters into $V_{CKM}$ to improve the fitting phenomenologically. \\

Considering the unitarity of $V_{CKM}$, several parameters $\alpha, ~\beta,~\gamma$ and $\alpha'$ are put into $V_{CKM}$ suitably as
\begin{eqnarray}
\vert V_{CKM} \vert = \left( \begin{array}{ccc}
r \sqrt{1+\gamma}          & s \sqrt{1+\beta}          & p \sqrt{1-\alpha} \\
s \sqrt{1-\beta}           & r \sqrt{1-\gamma}         & p \sqrt{1+\alpha} \\
p \sqrt{1-\alpha'}         & p \sqrt{1+\alpha'}        & q                   \end{array}\right).
\end{eqnarray}

Substituting the values in Eq.(16) into (17), one gets $q=0.99915$ directly and $p=0.29149$ with the unitary condition $\vert V_{ub} \vert^2
+ \vert V_{cb} \vert^2 + \vert V_{tb} \vert^2 =p^2 (1-\alpha)+p^2 (1+\alpha) +q^2=2 p^2 + q^2=1$.
Subsequently, $\alpha=0.98657$ is derived from the relation $\vert V_{cb} \vert^2 -\vert V_{ub} \vert^2 =2 p^2 \alpha$ and $\alpha^{\prime}=0.91070$ from $\vert V_{ts} \vert^2 -\vert V_{td} \vert^2 =2 p^2 \alpha^{\prime}$.
Similarly, $\vert V_{ud} \vert^2 +\vert V_{cs} \vert^2 =2 r^2$ gives $r=0.97393$ and $\vert V_{us} \vert^2 +\vert V_{cd} \vert^2 =2 s^2$ gives $s=0.22499$.
Thus, $\gamma=0.000852$ and $\beta =0.000622$ are also derived. \\

Substituting these parameters into Eq.(17), we will receive
\begin{eqnarray}
\vert V_{CKM} \vert  =
\left( \begin{array}{ccc}
 0.97434 & 0.22506  & 0.00338  \\
 0.22492 & 0.97351  & 0.04108  \\
 0.00871 & 0.04029  & 0.99915
 \end{array}\right).
\end{eqnarray}
It coincides the empirical values as good as to ${\bf O}(10^{-4})$ which is of the same order as experimental errors nowadays. \\

With these parameters, the next step is to test if there were in the parameters space of $x$, $y$, $x'$ and $y'$ sets of solutions satisfying the derived $p$, $q$, $r$ and $s$ values.
Before that, we may take advantage of $\vert V_{tb} \vert^2 ={{(x x'+y y' +x y x' y')^2}\over {(x^2+y^2+x^2 y^2)(x'^2+y'^2+x'^2 y'^2)}} = 0.99915^2$ to reduce one of the four parameters.
Numerically, a set of such parameters were found to satisfy the requirements with $x\cong 0.437$, $y\cong 2.994$, $x'\cong 0.396$ and $y' \cong 0.545$.
Accordingly, the phases defined in Eq.(11)-(14) can be predicted as: $\delta_1 \cong 9.9155^{o}$, $\delta_2 \cong 49.934^{o}$, $\delta_3 \cong 82.946^{o}$ and $\delta_4 \cong \delta_3 -\delta_1 \cong 73.030^{o}$.\\

Substituting the parameters $x$, $y$, $x'$ and $y'$ derived above and the quark masses $m_u=0.0022$GeV, $m_d =0.0049$GeV, $m_s=0.096$GeV, $m_c=1.275$GeV, $m_b=4.18$GeV and $m_t=173.21$GeV given in \cite{Patrignani2016}, the other parameters in Eq.(8) will also be given as
\begin{eqnarray}
A &=& 3.6713,  B=20.7777,  C=0.2526, \nonumber \\
A' &=& 1.3445, B'=1.7810,  C'=0.0139.
\end{eqnarray}

Consequently, the phases in $V_{CKM}$ can be rearranged as
\begin{eqnarray}
V_{CKM}  = \left( \begin{array}{ccc}
r \sqrt{1+\gamma}                      & s \sqrt{1+\beta}        & p \sqrt{1-\alpha} e^{i\delta} \\
s \sqrt{1-\beta}                       & r \sqrt{1-\gamma}       & p \sqrt{1+\alpha}\\
p \sqrt{1-\alpha'} e^{-i\delta}        & p \sqrt{1+\alpha'}      & q     \end{array}\right),
\end{eqnarray}
where $\delta = \delta_1-\delta_2 -2 \delta_3=-205.909^{o} $. \\

Though this phenomenological modification of CKM matrix improves the fitting between theoretical predictions and experimentally detected values almost perfectly.
However, the physical meaning of this fine-tuning is still obscure.
It is instinctive that these tiny deviations may be attributed to high order loop corrections and the best way to solve this puzzle is to calculate them directly to see if they were in agreement with these amonuts.
That will be a goal of our future studies. \\

\section{Application of the Hypothesis on the Standard Model}

In previous sections we studied the theoretical origin of CP violation in a model with two Higgs doublets.
However, after the derivation we find this hypothesis also applies to the Standard Model which has only one Higgs doublet.
If we have a way to derive a theoretical origin of CP violation in SM,
why shall we bother to employ an extra Higgs doublet to bring us the FCNC trouble?
In this section, we would like to demonstrate how the derivation can be applied to SM and even further to a 3HDM.   \\

The most general pattern of a $3 \times 3$ matrix is given in \cite{Lin2013} as
\begin{eqnarray}
M = \left( \begin{array}{ccc} A_1 +i D_1 & B_1+ i C_1 & B_2+ i C_2   \\
 B_4 + i C_4 & A_2 +i D_2 & B_3 + i C_3  \\ B_5 + i C_5 & B_6 + i C_6 & A_3 +i D_3 \end{array}\right),
\end{eqnarray}
where $A$, $B$, $C$ and $D$ parameters are all real. \\

However, this pattern is not only true for 2HDMs, it is always true for any $3 \times 3$ matrix.
If we extract a common factor $<\Phi>={v \over \sqrt{2}}$ out from the matrix in Eq.(21), where $\Phi$ is the unique Higgs doublet in SM,
Eq.(21) can be rewritten as
\begin{eqnarray}
M = <\Phi> {\bf Y} ={v \over \sqrt{2}} \left( \begin{array}{ccc} a_1 +i d_1 & b_1+ i c_1 & b_2+ i c_2   \\
    b_4 + i c_4 & a_2 +i d_2 & b_3 + i c_3  \\ b_5 + i c_5 & b_6 + i c_6 & a_3 +i d_3 \end{array}\right),
\end{eqnarray}
where $a_i  \equiv A_i {\sqrt{2}\over v}$, $b_i  \equiv B_i {\sqrt{2}\over v}$, $c_i  \equiv C_i {\sqrt{2}\over v}$ and $d_i  \equiv D_i {\sqrt{2}\over v}$ are elements of the Yukawa coupling matrix ${\bf Y}$.  \\

If we assume $M$ were a Hermitian matrix and spilt it into two real and imaginary components,
these two components will have the same patterns as those in Eq.(1) and they are respectively Hermitian, too.
Thus, the Hermitian condition $M_1 M_2^{\dagger}-M_2 M_1^{\dagger}=0$ and subsequent derivations in this manuscript all apply to them as well.
In this manuscript we not only solve the FCNC and CP problems in a 2HDM but also find the way to derive a theoretical origin of CP violation in the Standard Model. \\

Besides the application on SM, the $M$ matrix in Eq.(7) can be further split into three components corresponding to parameters $A$, $B$ and $C$ respectively as
\begin{eqnarray}
M = \left( \begin{array}{ccc}  A   & 0    & 0   \\  0    & A   &  0 \\  0    & 0    & A \end{array}\right)
  + \left( \begin{array}{ccc}   x B (y- {1 \over y})  & y B    & x B   \\   y B    & B ({y \over x}-{x \over y})  &  B \\  x B    & B    & 0 \end{array}\right) 
  +  i \left( \begin{array}{ccc}    0   & {C \over y} & -{C \over x}   \\    -{C \over y} & 0  & C \\     {C \over x} & C  & 0   \end{array}\right)  \nonumber \\
\end{eqnarray}
If we assign these three components to different Higgs doublets respectively, all of them will be diagonalized by the same $U$ matrix simultaneously.
That means even if we have three Higgs doublets, we can still annihilate the FCNC problem radically at tree-level. \\

\section{Conclusions and Discussions}

In this manuscript, the general CP-violating and FCNC-free pattern of quark-mass matrices in the 2HDM is derived as in Eq.(7) with the assumption of a Hermitian $M$ matrix.
Unlike our previous investigations, the mass eigenvalues and eigenvectors are derived directly in this manuscript and the $U$ matrix thus derived is found to depend on only two parameters.
Multiplying $U^{(u)}$ which depends only on parameters $x$ and $y$ and $U^{(d)\dagger}$ which depends only on parameters $x'$ and $y'$ together, a general CKM matrix is derived in Eq.(10) with elements given in Eq.(11)-(15).
The fitting between the theoretically derived CKM elements with corresponding empirical values are as good as ${\bf O}(10^{-2})$ which is orders improved than those given in our previous investigations ~\cite{Lin2013}. \\

Furthermore, a phenomenological way is imposed to improve the fitting by putting some fine-tuning parameters into the CKM matrix suitably.
That improves the fitting to as good as ${\bf O}(10^{-4})$ which is of the same order of present experimental errors.
However, the physical meaning of this fine-tuning is still obscure.
Theoretically the best way to examine it is to do loop calculations to see if they agree with these fine-tunings.
That will be an important goal of our future researches. \\

After the derivation of such a CP-violating and FCNC-free 2HDM, we find the derivations also apply to the Standard Model if the $M$ matrix is suitably rewritten as in Eq.(22).
Thus, the extra Higgs doublet which leads to the FCNC problem is in fact unnecessary for generating CP violations.  \\

Besides the 2HDM and SM, this pattern also applies to a model with three Higgs doublets.
As shown in Eq.(23), the $M$ matrix can be divided into three components which  correspond to parameters $A$, $B$ and $C$, respectively.
All these three components can be diagonalized by the same $U$ matrix simultaneously.
Thus, even if we have a 3HDM, there are still ways to forbid the appearance of FCNCs at tree-level completely. \\

In the future, examining the physical meaning of the fine-tunings with loop corrections is a main goal of our coming researches.
Besides, since during the process of the derivation we find the CP strength is not always the same, it hints a possibility that our universe may had experienced different CP strengths at different stages in its history.
That may hint an explanation of why the matter-antimatter imbalance we see nowadays cannot be accounted for by the presently detected CP strength.
That is also a goal of our future investigations. \\

\end{document}